\documentclass[aps,pre,showpacs,twocolumn,floats]{revtex4}
\usepackage{amssymb}

\usepackage{amsmath}
\usepackage{graphicx,psfig}
\usepackage{dcolumn}
\usepackage{bm}

\input{tcilatex}

\begin{document}

\title{Phonon superradiance and phonon laser effect in nanomagnets}
\author{E. M. Chudnovsky$^1$ and D. A. Garanin$^{1,2}$}
\affiliation{ \mbox{$^1$Department of Physics and Astronomy,
Lehman College, City University of New York,} \\ \mbox{250 Bedford
Park Boulevard West, Bronx, New York 10468-1589, U.S.A.} \\
\mbox{$^2$Institut f\"ur Physik, Johannes-Gutenberg-Universit\"at,
 D-55099 Mainz, Germany}}
\date{\today}

\begin{abstract}
We show that the theory of spin-phonon processes in paramagnetic solids must
take into account the coherent generation of phonons by the magnetic
centers. This effect should drastically enhance spin-phonon rates in
nanoscale paramagnets and in crystals of molecular nanomagnets.
\end{abstract}

\maketitle

The problem of spin-lattice interactions is almost as old as quantum theory
of solids \cite{AB}. It is about computing the amplitude of the transition
between spin states of a magnetic atom due to the spontaneous emission of a
phonon or due to, e.g., the Raman scattering of an existing thermal phonon.
These processes are responsible for the width of the paramagnetic resonance
and for other spin-lattice relaxation effects. In this Letter we shall
demonstrate that all previous works on the subject have missed the
collective nature of the relaxation when the wavelength of the phonon is
comparable to the distance between the magnetic atoms. General theoretical
treatment of this effect has become possible due to the recently established
mechanism of the dynamic spin-phonon coupling \cite{interaction},
\begin{equation}
\widehat{H}_{s-ph}=- \hbar \mathbf{S}\cdot \mathbf{\Omega ,\qquad \Omega(%
\mathbf{r) =}}\frac{1}{2}\nabla \times \mathbf{\dot{u}(\mathbf{r),}}
\label{HsphDef}
\end{equation}
where $\mathbf{S}$ is the spin of the atom, $\mathbf{\Omega }$ is the
angular velocity of the local rotation of the crystal due to a transverse
phonon, and $\mathbf{u}$ is the phonon displacement field. Eq.\ (\ref
{HsphDef}) has the following origin. The crystal field (i.e., the magnetic
anisotropy) that determines spin states of an atom in a solid is defined in
a local coordinate frame coupled to the crystal axes. In that coordinate
frame the local rotation of the lattice is equivalent to the magnetic field,
$\mathbf{B} = {\hbar} \mathbf{\Omega}/g {\mu}_{B}$ (with $g$ being the
gyromagnetic factor for the spin), that has Zeeman interaction with $\mathbf{%
S}$. Since the local rotations of the crystal lattice by phonons are small,
all matrix elements computed in the rotating frame and in the laboratory
frame are practically identical. The beauty of this approach is that it has
no unknown parameters. For, e.g., spontaneous emission of a phonon, all
parameters of the crystal field enter the interaction through the phonon
frequency only. The latter is determined by the distance between the spin
levels and can be directly measured, leaving no fitting parameters for
comparison between theory and experiment. Another way to look at this
interaction is through the conservation of the total angular momentum, spin
+ lattice. In order to conserve the angular momentum, the transition between
the spin states of an atom produces a local elastic twist that propagates
out as a transverse phonon. This Letter is based upon observation that an
assembly of the magnetic atoms in the excited state must interact via
transverse phonons very much the same as atoms interact via photons in the
laser physics. The deep analogy between the two problems can be seen by
noticing from Eq.\ (\ref{HsphDef}) that ${\dot{\mathbf{u}}}/2g{\mu}_{B}$ is
equivalent to the vector potential of the electromagnetic field. The huge $%
(c/v_t)^{3}$ ratio of the phonon density of states to the photon density of
states (with $v_t$ being the speed of the transverse sound) makes such a
phonon laser much more powerful than the photon laser based upon magnetic
dipoles.

We consider two nearly degenerate states of spin $\mathbf{S}$, described by
the Hamiltonian for pseudospin $1/2$,
\begin{equation}
\widehat{H}_{0}=-\frac{1}{2}\hbar \mathbf{\omega }_{0}\cdot \mathbf{\sigma
,\qquad }\hbar \mathbf{\omega }_{0}=W\mathbf{e}_{z}+\Delta \mathbf{e}_{x},
\label{HamHeffsigma}
\end{equation}
where $\mathbf{\sigma }$ is the Pauli matrix, $\Delta $ is the
splitting of the levels at the avoided crossing, and
$W=E_{-1}-E_{1}$ is the energy difference for the two degenerate
states at $\Delta =0$ ($\pm 1$ denoting spin up and down). We
label these states as $\left| \psi _{-1}\right\rangle $ and
$\left| \psi _{1}\right\rangle $, then $E_{\pm 1}=\left\langle
\psi _{\pm 1}\right| \widehat{H}_{\mathbf{S}}\left| \psi _{\pm
1}\right\rangle $ and $\Delta =2\left\langle \psi _{-1}\right|
\widehat{H}_{\mathbf{S}}\left| \psi _{1}\right\rangle ,$ where
$\widehat{H}_{\mathbf{S}}$ is the crystal-field Hamiltonian for
the spin $\mathbf{S}$. Choosing the new $z$ axis for pseudospins
in the direction of $\mathbf{\omega }_{0}$, one obtains the
eigenvalues and the eigenfunctions of $\widehat{H}_{0}$: $E_{\pm }=\pm (1/2)\sqrt{%
W^{2}+\Delta ^{2}}$ and $\left| \psi _{\pm }\right\rangle =\left. \left(
C_{\mp }\left| \psi _{1}\right\rangle \pm C_{\pm }\left| \psi
_{-1}\right\rangle \right) \right/ \sqrt{2}$ with $C_{\pm }=\sqrt{1\pm W/%
\sqrt{\Delta ^{2}+W^{2}}}$, whereas $\hbar \mathbf{\omega} _{0}=\sqrt{%
W^{2}+\Delta ^{2}}{\mathbf{e}}_z$. The projection of the spin-phonon
interaction, Eq.\ (\ref{HsphDef}), onto the two-state basis is
\begin{equation}
\widehat{H}_{s-ph}=-\frac{\hbar }{2}\mathbf{\sigma \cdot }%
\overleftrightarrow{\mathbf{g}}\mathbf{\cdot }\left( \nabla \times \mathbf{%
\dot{u}}\right) \equiv -\frac{\hbar }{2}\sigma _{\alpha }g_{\alpha \beta
}\left( \nabla \times \mathbf{\dot{u}}\right) _{\beta }\,.
\label{HamSpinPhononProjectedTensor}
\end{equation}
For tunneling between up and down states of spin $\mathbf{S}, $
$\;\left| \psi _{\pm 1}\right\rangle = \left| \pm S\right\rangle$
and
\begin{eqnarray}
\mathbf{g}_{x} &=&\mathbf{e}_{z}SC_{+}C_{-}=\mathbf{e}_{z}S\frac{\Delta }{%
\sqrt{W^{2}+\Delta ^{2}}},\qquad \mathbf{g}_{y}=\mathbf{0}  \notag \\
\mathbf{g}_{z} &=&-\mathbf{e}_{z}S\frac{1}{2}\left(
C_{+}^{2}-C_{-}^{2}\right) =-\mathbf{e}_{z}S\frac{W}{\sqrt{W^{2}+\Delta ^{2}}%
}.  \label{gConcreteFormSplitted}
\end{eqnarray}

For many two-state particles coupled to phonons the effective many-body
Hamiltonian is
\begin{equation}
\widehat{H}=-\frac{\hbar }{2}\sum_{i}\mathbf{\omega }_{0}\mathbf{\cdot
\sigma }_{i}-\frac{\hbar }{2}\sum_{i}\mathbf{\sigma }_{i}\mathbf{\cdot }%
\overleftrightarrow{\mathbf{g}}\mathbf{\cdot }\left[ \nabla \times \mathbf{%
\dot{u}(r}_{i})\right] +\widehat{H}_{ph}.  \label{HamManyParticles}
\end{equation}
where $\widehat{H}_{ph}$ is the Hamiltonian of non-interacting phonons. We
use canonical quantization of phonons,
\begin{equation}
\mathbf{u}=\sqrt{\frac{\hbar }{2MN}}\sum_{\mathbf{k}\lambda }\frac{\mathbf{e}%
_{\mathbf{k}\lambda }e^{i\mathbf{k\cdot r}}}{\sqrt{\omega _{k\lambda }}}%
\left( a_{\mathbf{k}\lambda }+a_{-\mathbf{k}\lambda }^{\dagger }\right) ,
\label{uQuantized}
\end{equation}
where $M$ is the mass of the unit cell, $N$ is the number of cells, $\mathbf{%
e}_{\mathbf{k}\lambda }$ are unit polarization vectors, $\lambda =t,t,l$
denotes polarization, and $\omega _{k\lambda }=v_{\lambda }k$ is the phonon
frequency. The spin-phonon Hamiltonian then becomes
\begin{equation}
\widehat{H}_{s-ph}=-\frac{\hbar }{2}\sum_{\mathbf{k}\lambda }\sum_{i}e^{i%
\mathbf{k\cdot r}_{i}}\mathbf{\sigma }_{i}\mathbf{\cdot G}_{\mathbf{k}%
\lambda }\left( a_{\mathbf{k}\lambda }-a_{-\mathbf{k}\lambda }^{\dagger
}\right) \,,  \label{HamSPk}
\end{equation}
where $\mathbf{G}_{\mathbf{k}\lambda }$ is due to transverse
phonons only, $\lambda =t,t$,
\begin{equation}
\mathbf{G}_{\mathbf{k}\lambda }\equiv \sqrt{\frac{\hbar \omega _{k\lambda }}{%
2MN}}\overleftrightarrow{\mathbf{g}}\mathbf{\cdot }\left[ \mathbf{k}\times
\mathbf{e}_{\mathbf{k}\lambda }\right] ,\qquad \mathbf{G}_{\mathbf{k}\lambda
}=-\mathbf{G}_{-\mathbf{k}\lambda }.  \label{GCouplingDef}
\end{equation}

In the Heisenberg representation for time-dependent operators, $i\hbar [d%
\widehat{O}(t)/dt] = [ \widehat{O}(t), \widehat{H}]$, one obtains, with
account of the commutation relations, $[ \sigma _{i\alpha },\sigma _{j\beta
}] = 2i{\epsilon}_{\alpha \beta \gamma }\sigma _{i\gamma }\delta _{ij}$ and $[ a_{%
\mathbf{k}\lambda },a_{\mathbf{k}^{\prime }\lambda ^{\prime
}}^{\dagger }] = \delta _{\mathbf{kk'}}\delta _{\lambda \lambda
^{\prime }}$, the following coupled equations:
\begin{eqnarray}
\dot{\mathbf{\sigma }}_{i}= \mathbf{\sigma }_{i}\times [ \mathbf{\omega }%
_{0}+\sum_{\mathbf{k}\lambda }e^{i\mathbf{k\cdot r}_{i}}\mathbf{G}_{\mathbf{k%
}\lambda }( a_{\mathbf{k}\lambda }-a_{-\mathbf{k}\lambda }^{\dagger } ) ]
\label{sigmaEq} \\
\dot{a}_{\mathbf{k}\lambda }=-\left( i\omega _{k\lambda }+\gamma _{k\lambda
}\right) a_{\mathbf{k}\lambda }+\frac{i}{2}\sum_{i}e^{-i\mathbf{k\cdot r}%
_{i}}\mathbf{\sigma }_{i}\mathbf{\cdot G}_{\mathbf{k}\lambda }\,.
\label{aEq}
\end{eqnarray}
Due to the linear interaction between $\mathbf{\sigma }_{i}$ and
$\mathbf{u}$, Eq.\ (\ref {HamManyParticles}) conserves
$(\sum_{i}\mathbf{\sigma }_{i})^2$ for uniform rotations.
Consequently, $\mathbf{\sigma }_{i}$ in Eqs.\ (\ref {sigmaEq}) and
(\ref {aEq}) can be treated as a semiclassical average over
distances large compared to the spacing between magnetic atoms but
small compared to the phonon wavelength. We are interested in the
interaction of such local polarization with a semiclassical phonon
field. One can write the Heisenberg operators as
\begin{equation}
{\sigma }_{i+}(t)={\tilde{\sigma}}_{i+}(t)e^{-i\omega
_{0}t},\qquad a_{\mathbf{k}\lambda
}(t)=\tilde{a}_{\mathbf{k}\lambda }(t)e^{-i\omega _{k\lambda
}t}\;,
\end{equation}
where ${\tilde{\sigma}}_{i+}(t)$ and $\tilde{a}_{\mathbf{k}\lambda
}(t) $ are slow variables. Keeping only resonant terms in Eqs.\
(\ref {sigmaEq}) and (\ref{aEq}), and neglecting the hybridization
of phonon and pseudospin modes, one obtains after integration on
time
\begin{equation}
a_{\mathbf{k}\lambda }=\frac{1}{4}\frac{\sum_{i}e^{-i\mathbf{k\cdot r}%
_{i}}\sigma _{i+}G_{\mathbf{k}\lambda ,-}}{\omega _{k\lambda }-\omega
_{0}-i\gamma _{k\lambda }} \;,
\end{equation}
where $G_{\mathbf{k}\lambda ,\pm }=G_{\mathbf{k}\lambda ,x}\pm iG_{\mathbf{k}
\lambda ,y}$. The insertion of $a_{\mathbf{k}\lambda }$ and $a_{-\mathbf{k}%
\lambda }^{\dagger }$ into Eq.\ (\ref{sigmaEq}) then gives the
system of equations:
\begin{eqnarray}
\dot{\sigma} _{i\pm } &=&\mp i\omega _{0}\sigma _{i\pm }-\frac{1}{2}\sigma
_{iz}\sum_{j}\Gamma _{ij}\sigma _{j\pm }  \notag \\
\dot{\sigma}_{iz} &=&\frac{1}{4}\sum_{j}\Gamma _{ij}\left( \sigma
_{j+}\sigma _{i-}+\sigma _{j-}\sigma _{i+}\right)
\label{sigmaEqplusminusFin}
\end{eqnarray}
with
\begin{equation}
\Gamma _{ij}=\frac{1}{N}\sum_{\mathbf{k}}e^{i\mathbf{k\cdot }\left( \mathbf{r%
}_{i}-\mathbf{r}_{j}\right) }V_{\mathbf{k}}^{2}\frac{2\gamma _{k}}{\left(
\omega _{k}-\omega _{0}\right) ^{2}+\gamma _{k}^{2}}  \label{GammaijRes}
\end{equation}
and
\begin{equation}
V_{\mathbf{k}}^{2}\equiv \frac{N}{4}\sum_{\lambda = t,t}\left( G_{\mathbf{k}%
\lambda ,x}^{2}+G_{\mathbf{k}\lambda ,y}^{2}\right) \;.
\label{V2Def}
\end{equation}

For a small sample of size, $L\lesssim \lambda _{0}\sim
v_{t}/\omega _{0}$, containing $\mathcal{N}\gg 1$ magnetic atoms
or molecules ($1\ll \mathcal{N}\ll N$) that are
initially polarized in one direction, one can use $e^{i\mathbf{k\cdot }%
\left( \mathbf{r}_{i}-\mathbf{r}_{j}\right) }\cong 1$ in Eq.\ (\ref
{GammaijRes}). Eqs.\ (\ref{sigmaEqplusminusFin}) then preserve the length of
the total polarization, $\sum_{j}\mathbf{\sigma }_{j}$. The unit vector $%
\mathbf{\sigma }=(1/\mathcal{N})(\sum_{j}\mathbf{\sigma }_{j})$ satisfies
the Landau-Lifhitz equation,
\begin{equation}
\dot{\mathbf{\sigma }}=\left[ \mathbf{\sigma \times \omega }_{0}\right]
-\alpha \left[ \mathbf{\sigma \times }\left[ \mathbf{\sigma \times \omega }%
_{0}\right] \right] \;,  \label{LLE}
\end{equation}
where $\alpha =\mathcal{N}{\Gamma _{1}}/({2\omega _{0}})$ is the
dimensionless
damping constant and $\Gamma _{1}=N^{-1}\sum_{\mathbf{k}}V_{\mathbf{k}%
}^{2}2\pi \delta \left( \omega _{k}-\omega _{0}\right) $ is the
one-atom spin-phonon rate. Replacing
$N^{-1}\sum_{\mathbf{k}}\ldots$ by $v_{0}\int
d^{3}k/(2\pi )^{3}\ldots $ ($v_{0}$ being the unit cell volume) and writing $%
V_{\mathbf{k}}^{2}$ with the help of Eqs.\ (\ref{gConcreteFormSplitted}), (%
\ref{GCouplingDef}), and (\ref{V2Def}) as
\begin{equation}
V_{\mathbf{k}}^{2}=\frac{S^{2}}{8\hbar }\frac{\Delta ^{2}}{Mv_{t}^{2}}\frac{%
\omega _{k}^{3}}{\omega _{0}^{2}}\left( 1-\frac{k_{z}^{2}}{k^{2}}\right) \,,
\label{V-angular}
\end{equation}
one obtains for the one-atom spin-phonon rate \cite{interaction}:
\begin{equation}
\Gamma _{1}=\frac{S^{2}}{12\pi \hbar }\frac{\Delta ^{2}}{Mv_{t}^{2}}\frac{%
\omega _{0}^{3}}{{\omega }_{D}^{3}}=\frac{S^{2}}{12\pi \hbar
}\frac{\Delta ^{2}{\omega }_{0}^{3}}{\rho v_{t}^{5}}\,,
\label{one-atom-Gamma}
\end{equation}
where ${\omega }_{D}\equiv v_{t}/v_{0}^{1/3}$ is the Debye
frequency for the tranverse phonons and $\rho =M/v_{0}$ is the
mass density of the crystal. It is easy to check that $\Gamma
_{\rm SR}=\mathcal{N}\Gamma _{1}$ is the rate of the relaxation of
$\sigma _{z}$ in Eq.\ (\ref{LLE}), that is, the longitudinal
relaxation rate. The transverse relaxation rate for $\sigma _{\pm
}$ is $\Gamma _{\rm SR}/2$. This is Dicke superradiance
\cite{Dicke} with
the rate exceeding the one-atom spin-phonon rate by a large factor $%
\mathcal{N}$. Eq.\ (\ref{LLE}), among other situations, describes the case
when all pseudospins are initially prepared in the excited state, ${\sigma }%
_{z}=-1$, and then relax collectively towards the ground state, ${\sigma }%
_{z}=1$, by emitting coherent phonons of frequency ${\omega }_{0}$. Because $%
V_{\mathbf{k}}^{2}$ of Eq.\ (\ref{V-angular}) is proportional to ${\sin }%
^{2}\theta =1-k_{z}^{2}/k^{2}$, the angular distrubution of the emitted
phonons is maximal in the direction perpendicular to the quantization axis $%
z $.

For large samples, $\lambda \lesssim L$, the phase of the emitted
phonons is no longer constant throughout the sample, and one
should also study inhomogeneous solutions of Eqs.\
(\ref{sigmaEqplusminusFin}). For that purpose it is convenient to
switch to the Fourier harmonics $\sigma
_{i\pm }=\mathcal{N}^{-1}\sum_{\mathbf{q}}e^{\pm i\mathbf{\ q\cdot r}%
_{i}}\sigma _{\mathbf{q}\pm }\;$\ and $\sigma _{iz}=\mathcal{N}^{-1}\sum_{%
\mathbf{q}}e^{i\mathbf{\ q\cdot r}_{i}}\sigma _{\mathbf{q}z}.$
When the paramagnetic body is a small part of a large solid
matrix, the phonon modes are practically continuous in the ${\bf
k}$-space while ${\bf q}$ is quantized. The first of Eqs.\
(\ref{sigmaEqplusminusFin}) then becomes
\begin{equation}
\dot{\sigma}_{\mathbf{q}\pm }=\mp i\omega _{0}\sigma _{\mathbf{q}\pm }-\frac{%
1}{2}\sum_{\mathbf{q}^{\prime \prime }}\Gamma (\mathbf{q}^{\prime
\prime })\sigma _{\lbrack \pm \left( \mathbf{q-q}^{\prime \prime
}\right) ],z}\sigma _{\mathbf{q}^{\prime \prime }\pm } \,,
\label{EqsigplusmunusFF2}
\end{equation}
where
\begin{equation}
\Gamma (\mathbf{q})=\frac{1}{N}\sum_{\mathbf{k}}V_{\mathbf{k}}^{2}\left| F_{%
\mathbf{k-q}}\right| ^{2}\frac{\gamma _{k}}{\left( \omega _{k}-\omega
_{0}\right) ^{2}+\gamma _{k}^{2}}\;,  \label{Gamma1qDef}
\end{equation}
and $F_{\mathbf{k}}=\mathcal{N}^{-1}\sum_{i}e^{i\mathbf{k\cdot
r}_{i}}\cong V^{-1}\int d\mathbf{r}e^{i\mathbf{k\cdot r}}.$ For a
rectangular body of dimensions $L_{\alpha}$ (${\alpha}= x,y,z$) it
gives
\begin{equation}
F_{\mathbf{k}}=\prod_{\alpha}\frac{\sin (k_{\alpha }L_{\alpha }/2)}{%
k_{\alpha }L_{\alpha }/2}  \label{FkBox}
\end{equation}
whereas for a sphere of radius $R$ one obtains $F_{\mathbf{k}}=f(x)\equiv
3x^{-3}(\sin x-x\cos x)$ with $x=kR.$

\begin{figure}
\unitlength1cm
\begin{picture}(11,6)
\centerline{\psfig{file=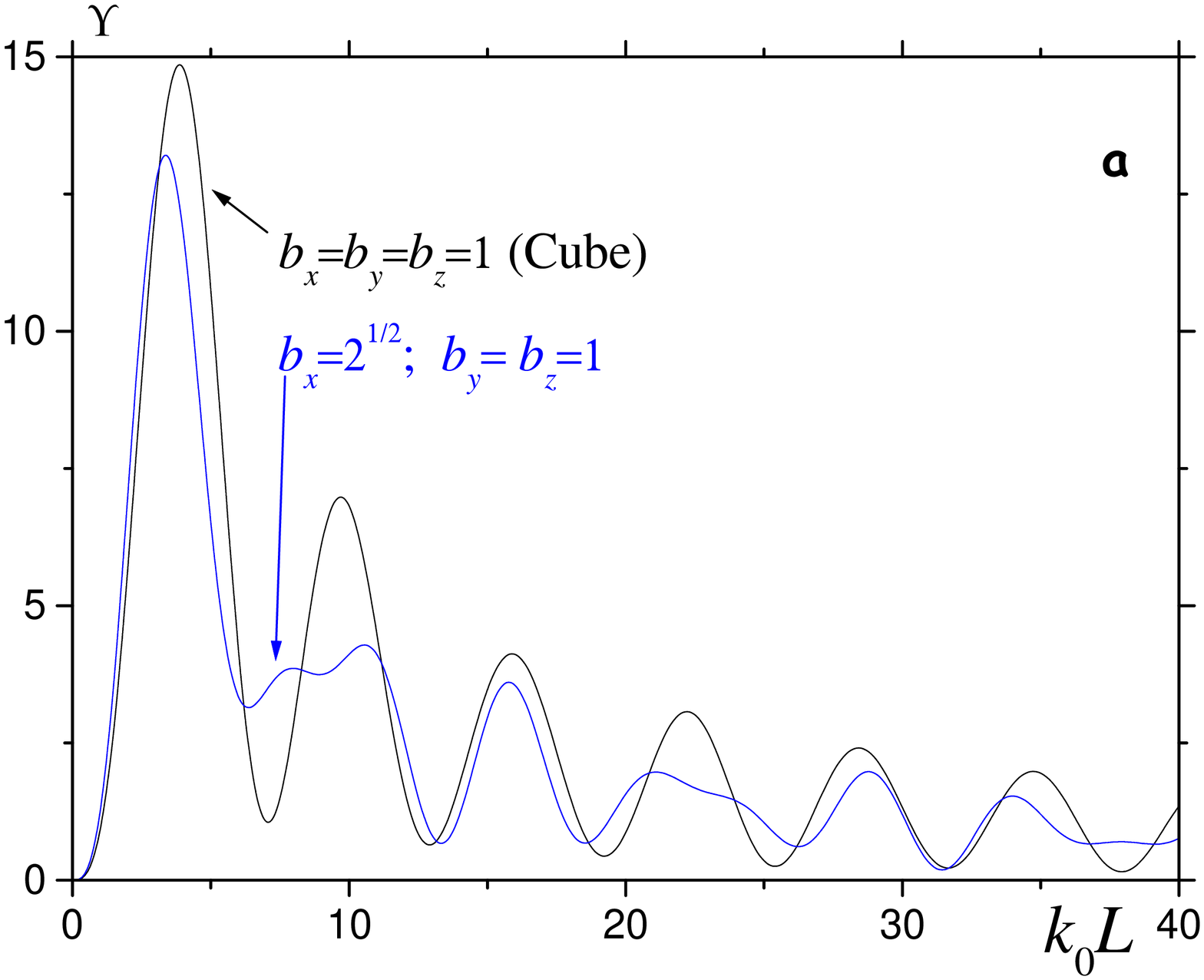,angle=-90,width=9cm}}
\end{picture}
\begin{picture}(11,6)
\centerline{\psfig{file=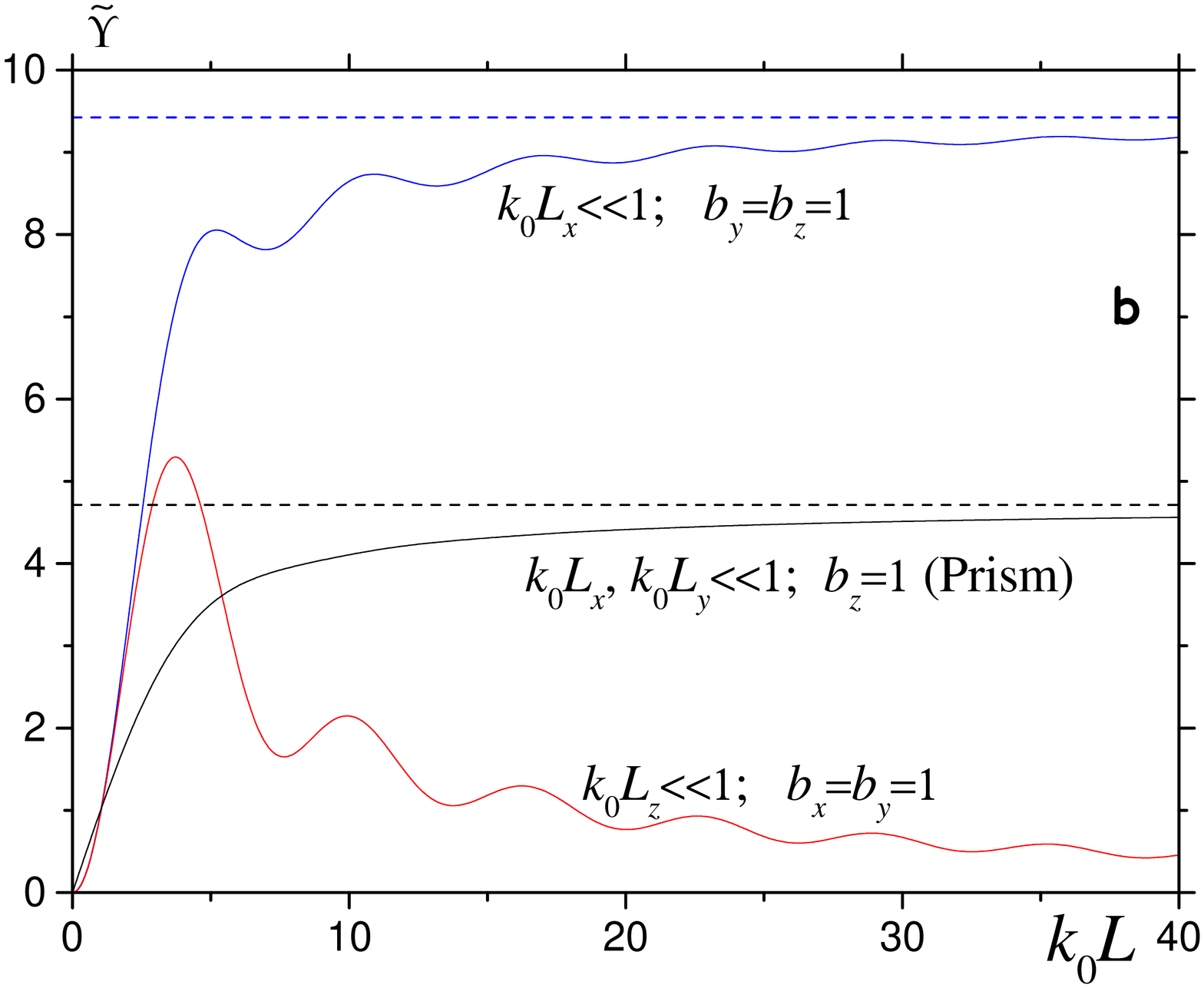,angle=-90,width=9cm}}
\end{picture}
\caption{\label{figY} $a$ -- $\Upsilon$ for rectangular geometry;
$b$ -- $\tilde \Upsilon$ for the film and prism geometries. $L_{\alpha }=b_{\alpha }L,$ $\alpha
=x,y,z.$
}
\end{figure}

Superradiance corresponds to the $\mathbf{q}=0$ mode in Eq.\ (\ref
{EqsigplusmunusFF2}). For ${\gamma }_{\mathbf{k}}\rightarrow 0$ it occurs at
a rate
\begin{eqnarray}
\Gamma _{\rm SR} &=&\mathcal{N}v_{0}\int \frac{d^{3}k}{\left( 2\pi \right) ^{3}}%
V_{\mathbf{k}}^{2}\left| F_{\mathbf{k}}\right| ^{2}2\pi \delta \left( \omega
_{k}-\omega _{0}\right)  \notag \\
&=&\mathcal{N}\Gamma _{1}\left. \left\langle V_{\mathbf{k}_{0}}^{2}\left| F_{%
\mathbf{k}_{0}}\right| ^{2}\right\rangle \right/ \left\langle V_{\mathbf{k}%
_{0}}^{2}\right\rangle ,  \label{Gamma1SRCrossover}
\end{eqnarray}
where $\left\langle \ldots \right\rangle $ denotes the average
over the directions of $\mathbf{k}_{0}$ that satisfy $\omega
_{k_{0}}=\omega _{0}$.
For $k_{\alpha }L_{\alpha }\ll 1$ one can replace $F_{\mathbf{k}_{0}}$ by $1$%
, and the second of Eqs.\ (\ref{Gamma1SRCrossover}) reduces to the
Dicke rate, $\Gamma _{\rm SR}=\mathcal{N}\Gamma _{1}$. In the
general case, the enhancement of the relaxation rate due to
superradiance is given by
\begin{equation}
\frac{\Gamma _{\rm SR}}{\Gamma
_{1}}=\frac{n_{M}}{k_{0}^{3}}\Upsilon , \qquad n_{M} =
\frac{\mathcal{N}}{V} \,.  \label{SREnhancementDef}
\end{equation}
where $n_M$ is concentration of magnetic atoms and
\begin{equation}
\Upsilon =k_{0}^{3}V\left. \left\langle V_{\mathbf{k}_{0}}^{2}\left| F_{%
\mathbf{k}_{0}}\right| ^{2}\right\rangle \right/ \left\langle V_{\mathbf{k}%
_{0}}^{2}\right\rangle  \label{YpsilonDef}
\end{equation}
depends on the wavelength and on the shape of the magnetic body.
Since the maximal value of $\Upsilon $ is of order $1$, the
maximal enhancement is always given by the dimensionless ratio
$n_{M}/k_{0}^{3}$. For a sphere one obtains $\Upsilon =\left( 4\pi
/3\right) \left( k_{0}R\right) ^{3}f^{2}(k_{0}R).$ The destructive
interference, $\Upsilon =0$, occurs at $ k_{0}R\cong \pi n,$ where
$n$ is an integer. At $k_{0}R\gg 1,$ $\;\Upsilon (k_{0}R)\sim
1/(k_{0}R)$. That is, ${\Gamma }_{\rm SR} $ goes down to ${\Gamma
}_{1}$ only at $k_{0}R\sim n_{M}/k_{0}^{3}$, that
can be very large. For a box of dimensions $%
L_{\alpha }=b_{\alpha }L,$ the geometrical factor $\Upsilon $ must
be computed numerically. For a cube of size $L,$ $\;\Upsilon
(k_{0}L)$
is qualitatively similar to that of the sphere. It almost vanishes at $%
k_{0}L=2\pi n$, which corresponds to the destructive interference of phonons
emitted perpendicular to the cube faces. The choice of $L_{x}\neq L_{y}$
causes a more complicated pattern shown in Fig.\ \ref{figY}a.

An interesting observation is that the increase of the $z$
dimension of the body beyond the wavelength of the radiation does
not lead to the decrease of the superradiant rate. The latter is
the consequence of the suppression of phonons emitted in that
direction, Eq.\ (\ref{V-angular}). For a sample of a prism shape
with $ k_{0}L_{x}\sim k_{0}L_{y}\ll 1$ and arbitrary $k_{0}L_{z}$,
one obtains $ \Upsilon =\left( k_{0}L_{x}\right) \left(
k_{0}L_{y}\right) \tilde{\Upsilon} \left(k_{0}L_{z} \right)$,
where $\tilde{\Upsilon}\left( u\right) \cong u$ at $u \equiv
k_{0}L_{z} \ll 1$ and $\tilde{\Upsilon}\left( u\right) \cong 3\pi
/2$ at $u\gg 1$, see Fig.\ \ref{figY}b. Another interesting case
is a flat sample with $k_{0}L_{x}\ll 1$, $\; b_{y}\sim b_{z}\sim
1$ and arbitrary $u=k_{0}L$. In this case one can write $\Upsilon
=\left( k_{0}L_{x}\right) \tilde{\Upsilon}\left(
u,b_{y},b_{z}\right)$, where $\tilde{\Upsilon} \cong
u^{2}b_{y}b_{z}$ at $u\ll 1$ and $\tilde{\Upsilon} \cong 3\pi $ at $u\gg 1$%
, Fig.\ \ref{figY}b. The latter is a thin-film limit for which
$\Upsilon$ depends neither on $b_{y,z}$ nor even on the shape of
the film area. The dependence of $\Upsilon$ on the thickness of
such $yz$-film is
\begin{equation}
\Upsilon =6\pi \frac{1-\cos \left( k_{0}L_{x}\right) }{k_{0}L_{x}}
\, .
\end{equation}
In this geometry the phonons are emitted in the direction $x$
perpendicular to the film. The emission rate can be as large as
$\Gamma _{\rm SR, \max} \approx 13.7 \left( n_{M}/k_{0}^{3}\right)
\Gamma _{1}$ at $k_{0}L_{x} \approx 2.33$, while the intensity of
sound is proportional to the film surface. One should notice the
difference from the case of the $xy$-film, Fig.\ \ref{figY}b, for
which the superradiant rate decreases with increasing the
dimensions of the film surface beyond the phonon wavelength.

%

Let us now consider the relaxation of the non-zero ${\sigma
}_{\mathbf{q}}$ harmonics in macroscopic samples for a particular
case of the maximal
inversion, ${\sigma }_{z}\cong -1$.  At $k_{0}L\gg 1$, because of the factor $\left| F_{%
\mathbf{k-q}}\right| ^{2}$ in Eq.\ (\ref{Gamma1qDef}), the phonon momenta $%
\mathbf{k}$ that contribute to Eq.\ (\ref{Gamma1qDef})  must be
very close to $\mathbf{q}$. Under the above conditions, Eq.\
(\ref{EqsigplusmunusFF2}) simplifies to
\begin{equation}
\dot{\sigma}_{\mathbf{q}\pm }=\mp i\omega _{0}\sigma _{\mathbf{q}\pm }+\frac{%
1}{2}\Gamma (\mathbf{q})\sigma _{\mathbf{q}\pm },  \label{laser-eq}
\end{equation}
where $\Gamma (\mathbf{q})$ is now the increment rate given by
\begin{equation}
\Gamma (\mathbf{q})=V_{\mathbf{q}}^{2}\frac{2\gamma _{q}}{\left(
\omega _{q}-\omega _{0}\right) ^{2}+\gamma _{q}^{2}} \,.
\label{Gamma1qLaser}
\end{equation}
The highest rate corresponds not to $\mathbf{q=0}$, as for the
superradiance, but to $\mathbf{q}$ that provides the minimal value
of $ \omega _{q}-\omega _{0}$ determined by the previously
neglected hybridization of the spin and phonon modes, $({\Delta
\omega })_{\min }\sim V_{\mathbf{k_{0}}}$. This is the laser mode
that grows at a rate
\begin{equation}
\Gamma _{L}=\frac{V_{\mathbf{k_{0},}\max }^{2}\gamma _{k_{0}}}{V_{\mathbf{%
k_{0},}\max }^{2}+\gamma _{k_{0}}^{2}}\,, \qquad
V_{\mathbf{k_{0},}\max }^{2}=\frac{S^{2}}{8\hbar }\frac{\Delta
^{2}\omega _{0}}{Mv_{t}^{2}}\,, \label{GammakResMax}
\end{equation}
where $V_{\mathbf{k_{0},}\max }^{2}$ is the maximal value of
$V_{\mathbf{ k_{0}}}^{2}$ achieved for $\mathbf{k}$ perpendicular
to the $z$ axis. The maximal value of $\Gamma _{L}$ on $\gamma
_{k_{0}}$ is ${\Gamma }_{L,\max }=V_{\mathbf{k_{0},}\max }$, and
the enhancement of the spin-phonon relaxation is very large:
\begin{equation}
\frac{{\Gamma }_{L,\max }}{{\Gamma }_{1}}=\frac{3\sqrt{2}}{S}\left( \frac{{%
\omega }_{D}}{{\omega }_{0}}\right) ^{3}\left( \frac{\hbar {\omega }_{0}}{%
\Delta }\right) ^{1/2}\left( \frac{Mv_{t}^{2}}{\Delta }\right)
^{1/2}. \label{laser-enhancement}
\end{equation}
At $V_{\mathbf{k_{0},}\max } \ll \gamma _{k_{0}}$, Eq.\
(\ref{GammakResMax}) reduces to the well-known laser formula,
$\Gamma _{L} = V_{\mathbf{k_{0},}\max }^{2}/\gamma _{k_{0}}$. As
in optical lasers, the increment rate ${\Gamma }_{L}$ must compete
with the dephasing rate ${\Gamma }_{2}$ due to, e.g.,
inhomogeneous broadening of $\Delta $. The laser generation
requires $V_{\mathbf{k_{0},\max }}^2>{\gamma }_{k_{0}}{\Gamma
}_{2}$. It will be the preferred decay mode of the inverted
population of spin states if ${ \omega}_{0}=\sqrt{W^{2}+\Delta
^{2}}/\hbar$ coincides with one of the frequencies of the
quantized phonon modes of the crystal. For a small crystal, this
condition can be satisfied only at certain values of $W$ and
$\Delta$. This can be achieved by continuously changing
longitudinal or transverse magnetic field. Since the distance
between these resonant field values is inversely proportional to
the size of the body, the resonance condition, at a finite
${\gamma}_{k_0}$, must be satisfied at any field for a body of
size $L \gtrsim v_t/{\gamma}_{k_0}$.

Our results on phonon superradiance alter the existing theory of
EPR in nanomagnets at low temperature. They show that one cannot
properly account for the width of the EPR by simply computing
one-phonon processes for isolated magnetic atoms. When the
wavelength of the phonons, ${\lambda}_0 = 2 \pi/k_0$, is greater
than the distance between magnetic atoms, the collective processes
come into play, greatly enhancing the rates. Observation of EPR
lines requires $k_{B}T \lesssim {\hbar}{\omega}_0$, which at $T
\lesssim 1\,$K translates into ${\lambda}_0 \lesssim 50\,$nm.
Thus, in the kelvin temperature range, our results on
superradiance are applicable to
nanomagnets of size below $50\,$nm and to thin films of thickness less than $%
50\,$nm. Larger samples should exhibit superradiance at lower temperatures.
The direct study of phonon superradiance requires inverted population of
spin states. In conventional paramagnets this can be achieved in a pulsed
magnetic field. In microscrystals of molecular nanomagnets the inverted
population of spin states is easy to achieve even in a slowly varying field
\cite{Sessoli}. Note that in crystals of small size the phonon superradiance
always wins over the electromagnetic superradiance \cite{CG-02}.

The existence of the laser phonon mode may be consistent with the
experimental evidence \cite{PRL-96,Friedman,Dalal,Dressel} that
the low-temperature spin-phonon rates in molecular nanomagnets are
few orders of magnitude higher than the computed one-molecule
spin-phonon rates \cite {Villain,GC-97}. In molecular magnets the
two-level system is formed every time the external magnetic field,
$B$, produces a resonance between the states described by the
magnetic quantum numbers $m$ and $m^{\prime}$. For the $(m,m')$
resonance, $S$ in all our formulas must be replaced by
$|m-m^{\prime}|/2$, while $W=|m-m^{\prime}|g{\mu}_{B}B$. Consider,
e.g., a typical field-sweep experiment in Fe$_8$, when the field
crosses the $(10,-8)$ resonance and is $0.1$T past the resonance.
Taking $\Delta \simeq 10^{-4}$K \cite{WS}, $M \simeq 5 \times
10^{-21}$g, and $v_t \simeq 10^{5}$cm/s, one obtains: ${\omega}_0
\simeq 10^{11}$s$^{-1}$, $\; {\Gamma}_1 \simeq 10^{-5}$s$^{-1}$,
but ${\Gamma}_{L,\max} \simeq 10^4$s$^{-1}$, which would provide
the $m=10 \Rightarrow m^{\prime}=-8$ spin-phonon relaxation in
less than $1$-ms time. This may explain spin-phonon avalanches
observed in sufficiently large crystals of molecular magnets \cite
{Paulsen,Barco}. The laser rate $\Gamma _{L} =
V_{\mathbf{k_{0},}\max }^{2}/\gamma _{k_{0}}$ has a distinctive
${\omega}_0$ dependence that should be seen in experiment. At low
temperature $\gamma _{k_{0}} \propto {\omega}_{0}^{4}$ due to the
scattering of phonons by the inhomogeneities of the crystal
\cite{Garanin}, while $V_{\mathbf{k_{0},}\max }^{2} \propto
{\omega}_0$. Consequently, $\Gamma _{L} \propto {\omega}_0^{-3}$,
which is opposite to the ${\omega}_0$ dependence of the
one-molecule spin-phonon rate, $\Gamma _1 \propto
{\omega}_{0}^{3}$.

This work has been supported by the NSF Grant No. EIA-0310517.

\end{document}